\documentclass[a4paper,11pt]{article}
\usepackage{pos}

\title{Lee-Yang zeros in heavy-quark QCD}

\author*[a,b]{Masakiyo Kitazawa}
\author[a]{Tatsuya Wada}
\author[c]{Kazuyuki Kanaya}

\affiliation[a]{Yukawa Institute for Theoretical Physics,
Kyoto University, Kyoto 606-8502, Japan}
\affiliation[b]{
  J-PARC Branch, KEK Theory Center, 
  Institute of Particle and Nuclear Studies, KEK, \\ Tokai, Ibaraki 319-1106, Japan}
\affiliation[c]{Tomonaga Center for the History of the Universe, University of Tsukuba, \\ Tsukuba, Ibaraki 305-8571, Japan}

\emailAdd{kitazawa@yukawa.kyoto-u.ac.jp}
\emailAdd{tatsuya.wada@yukawa.kyoto-u.ac.jp}
\emailAdd{kanaya@ccs.tsukuba.ac.jp}

\abstract{We explore the distribution of Lee-Yang zeros around the critical point that appears in the heavy-quark region of QCD at nonzero temperature in lattice numerical simulations. With the aid of the hopping-parameter expansion that is well justified around the critical point in our setting, our numerical analysis is capable of analyzing the partition function for complex parameters with high accuracy. This enables precise analyses of the Lee-Yang zeros around the critical point. We study their finite-size scaling around the critical point. We also propose new methods to utilize the scaling behavior of the Lee-Yang zeros for fixing the location of the critical point.}

\FullConference{The XVIth Quark Confinement and the Hadron Spectrum Conference (QCHSC24)\\
 19-24 August, 2024\\
 Cairns Convention Centre, Cairns, Queensland, Australia\\}

\begin{document}
\begin{flushright}
    YITP-25-47, J-PARC-TH-0314
\end{flushright}
\maketitle

\section{Introduction}

An interesting feature of the medium described by quantum chromodynamics (QCD) is that the appearance of critical points (CPs) is ubiquitously expected on phase diagrams of various parameters controlling the system. 
Among them, CPs that appear when the quark masses are varied from real values have attracted much attention over the last decades~\cite{Jin:2017jjp,Ejiri:2019csa,Kuramashi:2020meg,Cuteri:2020yke,Philipsen:2021qji,Kiyohara:2021smr,Aarts:2023vsf,Ding:2024sux,Ashikawa:2024njc},
since they manifest themselves at zero baryon chemical potential where lattice-QCD Monte Carlo simulations can be carried out without the sign problem.
However, the lattice simulations of the CPs are still difficult both in the light- and heavy-quark regions. For the light-quark region, even the existence of a CP is still controversial~\cite{Kuramashi:2020meg,Philipsen:2021qji,Ding:2024sux}. For the heavy-quark region, while the existence of the CP is well established, it is known that large-volume simulations are needed for proper reproduction of the expected scaling behavior~\cite{Cuteri:2020yke,Kiyohara:2021smr,Ashikawa:2024njc}. 

In this proceeding, we study the CP in the heavy-quark region using Lee-Yang zeros (LYZ)~\cite{Lee:1952ig}. Recently, there have been attempts to locate the QCD CP at nonzero baryon chemical potential with the use of the LYZ and Lee-Yang edge singularity~\cite{Basar:2023nkp,Clarke:2024ugt,Schmidt:2025ftp,Adam:2025pii}. Motivated by these studies, we explore the behavior of the LYZ around the CP in heavy-quark QCD focusing on their finite-size scaling. We show that the ratio of LYZ (LYZR) has useful properties for studying CPs in general systems, and propose their use in numerical simulations~\cite{Wada:2024qsk}. 
This method is then adopted to the CP in heavy-quark QCD. By comparing these results with the conventional analysis based on the Binder cumulant~\cite{Binder:1981sa}, we show that the LYZR method can determine the location of the CP successfully.

\section{Lee-Yang-zero ratio (LYZR) method}
\label{sec:LYZR}

\subsection{LYZR in Ising model}
\label{sec:Ising}

To illustrate the LYZR method~\cite{Wada:2024qsk}, we begin with the Ising model of the Hamiltonian
\begin{align}
    H(h)=-J\sum_{\langle ij \rangle} s_is_j-h\sum_{i} s_i, \label{eq:def_Ising_Hamiltonian}
\end{align}
on the cubic lattice of size $L^3$.
This model has a critical point at $T=T_c$ with $T_c/J\simeq4.5115$~\cite{Ferrenberg:2018zst} and $h=0$, and a first-order phase transition at $h=0$ for $t\equiv(T-T_c)/T_c<0$. In the following, we set $J=1$.

We denote the partition function of this model as $Z(t,h,L^{-1})$, which satisfies $Z(t,h,L^{-1})=Z(t,-h,L^{-1})$ from the symmetry.
The LYZ in the Ising model are defined as the values of $h\in\mathbb{C}$ satisfying $Z(t,h,L^{-1})=0$ for $t\in\mathbb{R}$~\cite{Lee:1952ig}. It is known that the LYZ in this model for finite $L$ appear only on the pure-imaginary axis~\cite{Lee:1952ig}. In what follows, we denote the LYZ with ${\rm Im}h>0$ as $h=h_{\rm LY}^{(n)}(t,L)$, where $n=1,2,\cdots$ represents different LYZ for ${\rm Im}h>0$ so that $0<{\rm Im} h_{\rm LY}^{(1)}(t,L)<{\rm Im} h_{\rm LY}^{(2)}(t,L)<\cdots$. We notice that $h_{\rm LY}^{(n)}(t,L)$ should be regular functions of $t$ for finite $L$, because $Z(t,h,L^{-1})$ is a regular function of $t$ and $h$.

From the finite-size scaling (FSS), dependence of $Z(t,h,L^{-1})$ on $t$ and $h$ near the CP is represented by the scaling function $\tilde{Z}$ as 
\begin{align}
    Z(t,h,L^{-1}) = \tilde{Z}(L^{y_t}t,L^{y_h}h) ,
\label{eq:FSS}
\end{align}
for sufficiently large $L$ with 
the scaling dimensions $y_t\simeq 1.588$ and $y_h\simeq 2.482$ for the three-dimensional Ising model~\cite{Ferrenberg:2018zst}. 
From Eq.~\eqref{eq:FSS}, one finds
\begin{align}
    h_{\rm LY}^{(n)}(t,L) = L^{-y_h} \, \tilde{h}_{\rm LY}^{(n)}(L^{y_t}t) ,
    \label{eq:tildeh_LY}
\end{align}
with $\tilde{h}_{\rm LY}^{(n)}(\tilde t)$ satisfying $\tilde Z(\tilde t,\tilde{h}_{\rm LY}^{(n)}(\tilde t))=0$.

\begin{figure}
    \centering
\includegraphics[width=0.25\textwidth]{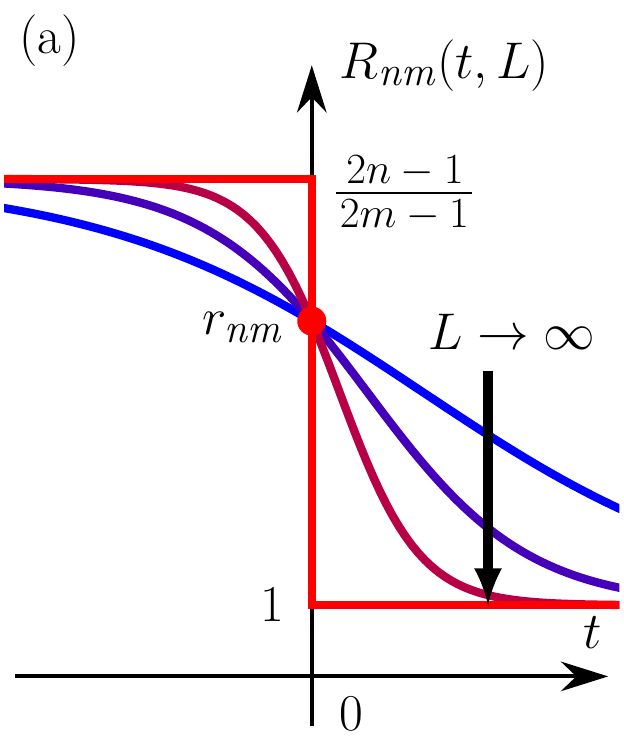}
\hspace{0.15\textwidth}
\includegraphics[width=0.25\textwidth]{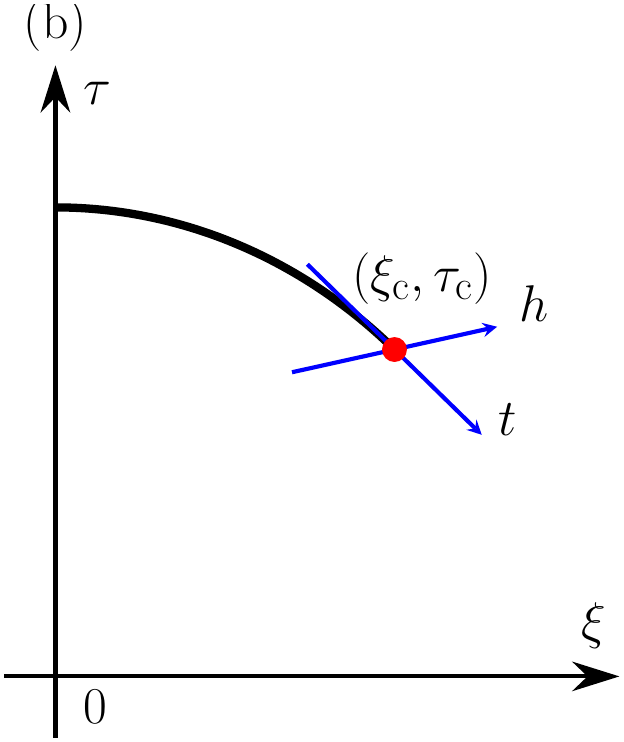}
    \caption{(a) Behavior of Lee-Yang zero ratio $R_{nm}(t,L)$ as a function of $t$~\cite{Wada:2024qsk}. Curved lines represent $R_{nm}(t,L)$ at finite $L$'s, while the red line represents $R_{nm}(t,L)$ in the limit $L\to\infty$.
    (b) Schematic phase diagram having a first-order phase-transition line (black line) and a CP (red circle). The Ising variables $t$ and $h$ encoded through the mapping of the scaling function are shown by the blue arrows.}
    \label{fig:cartoon}
\end{figure}

Now, let us focus on the ratio of two LYZ (LYZR) at the same $L$,
\begin{align}
    R_{nm}(t,L) = \frac{h_{\rm LY}^{(n)}(t,L)}{h_{\rm LY}^{(m)}(t,L)}
    = \frac{\tilde h_{\rm LY}^{(n)}(L^{y_t}t)}{\tilde h_{\rm LY}^{(m)}(L^{y_t}t)} .
    \label{eq:Rnm}
\end{align}
Let us first investigate the limiting behavior of Eq.~\eqref{eq:Rnm} for $L\to\infty$.
For $t<0$, corresponding to the discontinuity of the first-order phase transition at $h=0$, the LYZ are aligned with an equal distance as
\begin{align}
    h_{\rm LY}^{(n)}(t,L)=\frac{a(t)(2n-1)}{L^3} \; \xrightarrow[L\to\infty]{} \; 0 
    \quad
    (t<0,\;n:\textrm{finite}),
    \label{eq:LYES0}
\end{align} 
for sufficiently large $L$ with $a(t)$ being a pure-imaginary function~\cite{Ejiri:2005ts}. This means that the LYZR behave as $R_{nm}(t)=(2n-1)/(2m-1)$ for $t<0$.
For $t>0$, on the other hand, because of the regularity of $Z(t,h,L^{-1})$ at $h=0$, the distributions of LYZ for $L\to\infty$ must terminate at pure-imaginary points $h=\pm h_{\rm LYES}(t)$, which are called the Lee-Yang edge singularity. One thus finds that $R_{nm}(t)$ approaches unity for $L\to\infty$. 
\begin{align}
    h_{\rm LY}^{(n)}(t,L) \;
    \xrightarrow[L\to\infty]{} \; h_{\rm LYES}(t) 
    \quad
    (t>0,\;n:\textrm{finite}).
    \label{eq:LYES}
\end{align} 
To summarize, we find that the LYZR in the $L\to\infty$ limit behave as~\cite{Wada:2024qsk},
\begin{align}
    R_{nm}(t) \xrightarrow[L\to\infty]{}
    \begin{cases}
        \frac{2n-1}{2m-1} & (t<0) \\
        1 & (t>0)
    \end{cases}
    \qquad
    (\mbox{finite}~ n,m).
    \label{eq:Rlim}
\end{align}
This behavior is depicted by the red line in Fig.~\ref{fig:cartoon}~(a). The rapid change at $t=0$ would be useful in determining a CP in numerical simulations. 

Next, to gain insights into the behavior of LYZR around $t=0$, we Taylor expand $\tilde{h}_{\rm LY}^{(n)}(\tilde t)$ at $\tilde t=0$ as 
\begin{align}
    \tilde{h}_{\rm LY}^{(n)}(\tilde t) = i\big( X_{n} + Y_{n}\tilde t + {\cal O}(\tilde{t}\,^2) \big) ,
    \label{eq:XY} 
\end{align}
with real coefficients $X_n$ and $Y_n$.
Substituting Eq.~\eqref{eq:XY} into Eq.~\eqref{eq:Rnm}, one obtains 
\begin{align}
    & R_{nm}(t,L) = r_{nm} + c_{nm} L^{y_t} t + {\cal O}(t^2) ,
    \label{eq:Rlinear}
\end{align}
with $r_{nm} = X_n/X_m$ and $c_{nm} = r_{nm}\,(Y_n/X_n - Y_m/X_m )$. From this result one finds that $R_{nm}(t,L)=r_{nm}$ is independent of $L$ at $t=0$, while the slope scales as $L^{y_t}$. This means that the LYZRs for different $L$ intersect at the CP at $t=0$, as schematically shown in Fig.~\ref{fig:cartoon}~(a). We notice that these properties of the LYZR are similar to the Binder cumulants~\cite{Binder:1981sa,Ashikawa:2024njc}. They thus provide us with alternative methods to determine the CP from the intersection point in numerical simulations. 

\subsection{LYZR in general systems}
\label{sec:general}

Next, we extend the above argument to general systems having a CP that belongs to the same universality class as the Ising model. We suppose that the partition partition function of the system, ${\cal Z}(\tau,\xi,L^{-1})$, is described by variables $\tau$ and $\xi$ with the system size $L$. We also assume that the phase diagram of this system on the $\tau$--$\xi$ plane has a first-order phase transition that terminates at a CP at $(\tau,\xi)=(\tau_{\rm c},\xi_{\rm c})$, as schematically 
shown in Fig.~\ref{fig:cartoon}~(b).
From the scaling hypothesis, the partition function in the vicinity of the CP is related to that of the Ising model as 
\begin{align}
    {\cal Z}(\tau,\xi,L^{-1}) = Z(\check t(\tau,\xi),\check h(\tau,\xi),L^{-1}) ,
    \label{eq:F=F}
\end{align}
for sufficiently large $L$, where $\check t(\tau,\xi)$ and $\check h(\tau,\xi)$ obey the linear relation
\begin{align}
    \begin{pmatrix}
        \check t \\ \check h
    \end{pmatrix}
    =
    \begin{pmatrix}
        a_{11} & a_{12} \\
        a_{21} & a_{22} 
    \end{pmatrix}
    \begin{pmatrix}
        \tau-\tau_{\rm c} \\ \xi-\xi_{\rm c}
    \end{pmatrix}
    \equiv
    A
    \begin{pmatrix}
        \delta\tau \\ \delta\xi
    \end{pmatrix} .
    \label{eq:A}
\end{align}

We define the LYZ in this system as zero points of ${\cal Z}(\tau,\xi,L^{-1})$ for $\xi\in\mathbb{C}$ and $\tau\in\mathbb{R}$, and denote the LYZ for ${\rm Im}\xi>0$ as $\xi=\xi_{\rm LY}^{(n)}(\tau,L)$, i.e. ${\cal Z}(\tau,\xi_{\rm LY}^{(n)}(\tau,L),L^{-1})=0$. Then, Eq.~\eqref{eq:F=F} leads to
\begin{align}
    L^{y_h} \check h\big(\tau,\xi_{\rm LY}^{(n)}(\tau,L)\big) 
    = \tilde{h}_{\rm LY}^{(n)}\big(L^{y_t}\check t(\tau, \xi_{\rm LY}^{(n)}(\tau,L)) \big),
    \label{eq:hxi}
\end{align}
and together with Eqs.~\eqref{eq:A} and~\eqref{eq:XY} one finds
\begin{align}
    L^{y_h} \big( a_{21} \delta\tau + a_{22} (\xi_{\rm LY}^{(n)}(\tau,L)-\xi_{\rm c}) \big) 
    = i\big( X_n + Y_n L^{y_t} \big( a_{11} \delta\tau + a_{12} (\xi_{\rm LY}^{(n)}(\tau,L)-\xi_{\rm c}) \big) \big)+ {\cal O}(\delta\tau^2),
\end{align}
which yields~\cite{Wada:2024qsk}
\begin{align}
    \xi_{\rm LY}^{(n)}(\tau,L)  
    = \xi_{\rm c} + \frac{ iX_n - ( a_{21} L^{y_h} -iY_n a_{11} L^{y_t} )\delta\tau}{ a_{22} L^{y_h} -iY_n a_{12}L^{y_t} }.
    \label{eq:xi_LY}
\end{align}
Expanding Eq.~\eqref{eq:xi_LY} by $L^{-1}$ one obtains 
\begin{align}
    &{\rm Re}\xi_{\rm LY}^{(n)}(\tau,L)  
    = \xi_{\rm c} - \frac{a_{21}}{a_{22}}\delta\tau + {\cal O}(L^{2\bar y}),
    \label{eq:Rexi} 
    \\
    &{\rm Im}\xi_{\rm LY}^{(n)}(\tau,L)  
    = \frac{X_n}{a_{22}} L^{-y_h} + \frac{Y_n\det A}{a_{22}^2} L^{\bar y} \delta\tau + {\cal O}(L^{2\bar y}),
    \label{eq:Imxi}
\end{align}
with $\bar y=y_t-y_h<0$. 

To introduce the LYZR in this system, we consider the ratios between the imaginary parts of $\xi_{\rm LY}^{(n)}$. By expanding them by $\delta\tau$ and $L^{-1}$ one obtains~\cite{Wada:2024qsk}
\begin{align}
    {\cal R}_{nm}(\tau,L) 
    = \frac{{\rm Im}\xi_{\rm LY}^{(n)}(\tau,L)}{{\rm Im}\xi_{\rm LY}^{(m)}(\tau,L)}
    = \big( r_{nm} + C_{nm} L^{y_t} \delta\tau + {\cal O}(\delta\tau^2) \big) 
    \big( 1 + D_{nm} L^{2\bar y} + {\cal O}(L^{4\bar y})\big),
    \label{eq:Rlinear2}
\end{align}
where $C_{nm}=c_{nm}\det A /a_{22}$ and $D_{nm}= -(Y_n^2-Y_m^2)a_{12}^2/a_{22}^2$. For $L\to\infty$, Eq.~\eqref{eq:Rlinear2} is dominated by the first bracket on the far-right-hand side. This means that the intersection point of Eq.~\eqref{eq:Rlinear2} converges to the CP as in the Ising model for $L\to\infty$. Notice that $r_{nm}=\lim_{L\to\infty}{\cal R}_{nm}(\tau_{\rm c},L)$ is the same as Eq.~\eqref{eq:Rlinear}, i.e. the value of ${\cal R}_{nm}(\tau,L)$ at the intersection point is unique in individual universality class. 

Now, let us compare Eq.~\eqref{eq:Rlinear2} with the behavior of the Binder cumulant~\cite{Binder:1981sa}
\begin{align}
    {\cal B}_4(\tau,L)={\rm min}_\xi \Bigg[\frac{\partial^4 {\cal F}(\tau,\xi,L^{-1})/\partial \xi^4)}{(\partial^2 {\cal F}(\tau,\xi,L^{-1})/\partial \xi^2)^2}\Bigg]+3
    \label{eq:B4}
\end{align}
with the free energy ${\cal F}$. From Eq.~\eqref{eq:B4} one obtains~\cite{Jin:2017jjp,Cuteri:2020yke}
\begin{align}
    {\cal B}_4(\tau,L) 
    = \big( b_4 + c_4 L^{y_t} \delta\tau + {\cal O}(\delta\tau^2) \big) 
    \big( 1 + d_4 L^{\bar y} + {\cal O}(L^{2\bar y})\big),
    \label{eq:Rlinear3}
\end{align}
where $d_4$ is proportional to $a_{12}$. Comparing this result with Eq.~\eqref{eq:Rlinear2}, one finds that the second bracket in Eq.~\eqref{eq:Rlinear3} converges slower than that in Eq.~\eqref{eq:Rlinear2} for $L\to\infty$. This implies that the finite-volume effect from $a_{12}\ne0$ is suppressed more quickly for $L\to\infty$ in ${\cal R}_{nm}(\tau,L)$ than ${\cal B}_4(\tau,L)$, which would be an advantage of the LYZR method.

\section{Heavy-Quark QCD}

We now apply the LYZR method to the analysis of the CP in the heavy-quark region. For simplicity, we focus on the case of degenerate $2$-flavor ($N_{\rm f}=2$) QCD. Our lattice QCD action is given by $S=S_{\rm g}- N_{\rm f}\ln\det M(\kappa)$ with the gauge action $S_{\rm g}$, the quark matrix $M(\kappa)$, and the hopping parameter $\kappa$ that is related to the bare quark mass $m_q$ as $\kappa=1/(2am_q+4)$. We use the plaquette action for $S_{\rm g}$. 

As the heavy-quark limit corresponds to $\kappa\to0$, in heavy-quark QCD it is useful to expand the quark determinant with respect to $\kappa$. This expansion is called the hopping parameter expansion (HPE). We denote the HPE as 
\begin{align}
    S = S_{\rm g} + S_{\rm LO} + S_{\rm NLO} + \cdots,
\end{align}
where $S_{\rm LO}$ and $S_{\rm NLO}$ are the leading and next-to-leading order terms of the HPE. We define $S_{\rm LO}$ to include the leading-order terms with and without the windings along the imaginary-time direction, and the same for higher orders. For $N_t=4$, for example,  $S_{\rm LO}$ is given by~\cite{Kiyohara:2021smr,Ashikawa:2024njc}
\begin{align}
    S_{\textrm{LO}} 
    = - 2 N_f N_c \kappa^4 (48N_\textrm{t}N_\textrm{s}^3\hat{P}+32N_\textrm{s}^3\hat{\Omega}_R),
    \label{eq:LO_HQ_action}
\end{align}
where $\hat{P}$ and $\hat{\Omega}$ are the plaquette and Polyakov-loop operators normalized such that they become unity in the weak-coupling limit and $N_{\rm c}=3$.

\begin{figure}
    \centering
\includegraphics[width=0.4\linewidth]{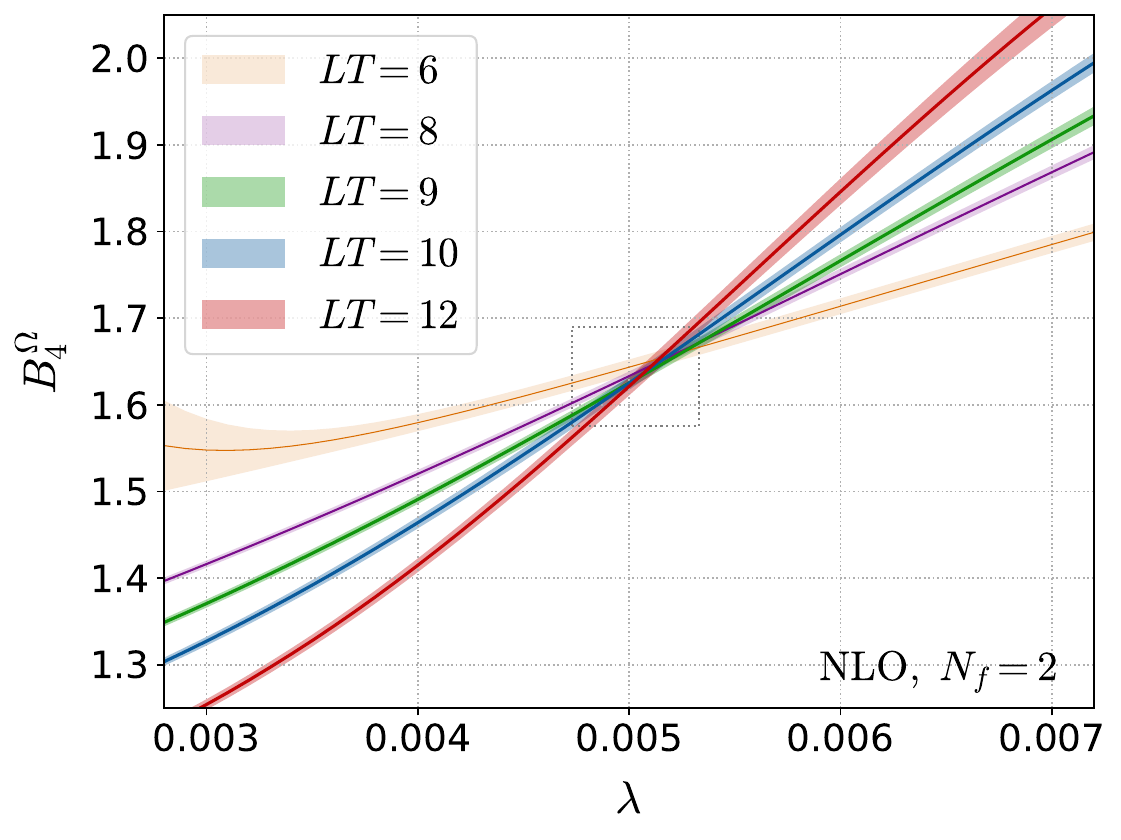}
\includegraphics[width=0.4\linewidth]{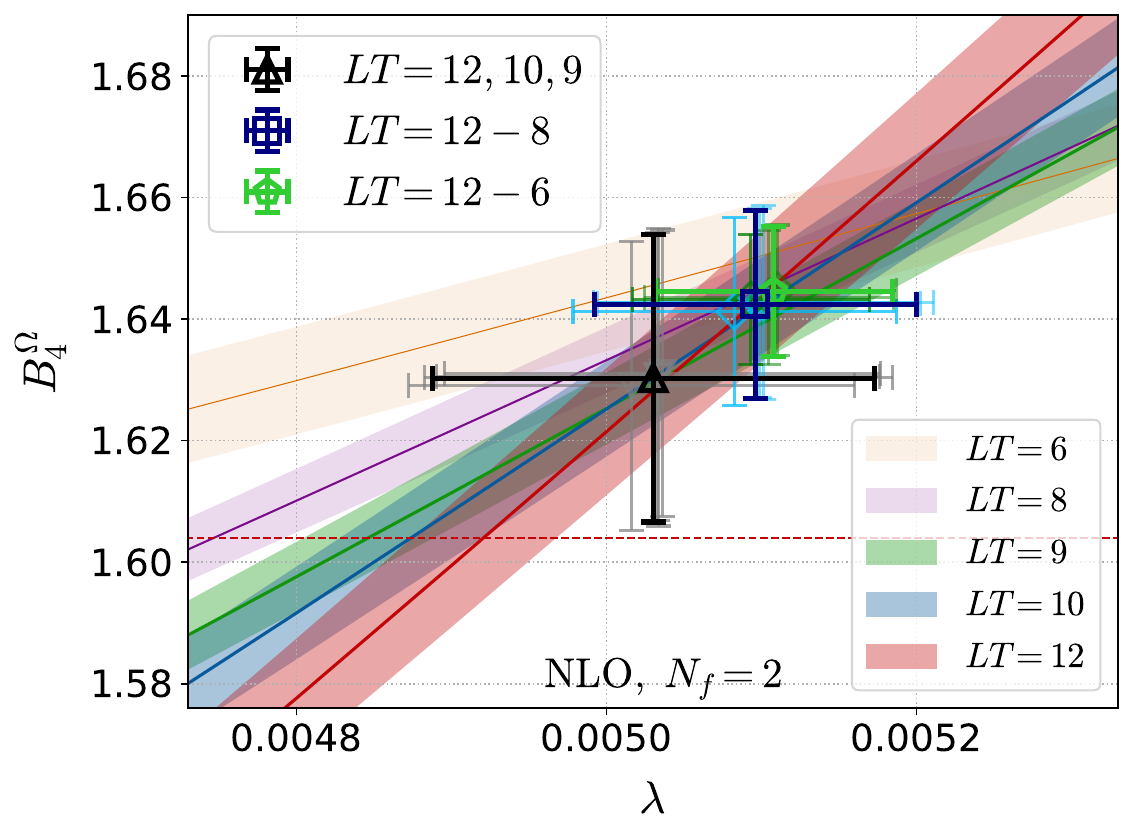}
    \caption{Binder cumulant of the Polyakov loop around the CP in heavy-quark QCD as a function of $\lambda= 2^{N_t+2} N_{\rm f} N_{\rm c} \kappa^{N_t}$ for various aspect ratios $LT=N_s/N_t$~\cite{Kiyohara:2021smr}. The right panel is an expansion of the left panel. The fit results of the intersection point are shown by markers with error bars.}
    \label{fig:B4}
\end{figure}

In Refs.~\cite{Kiyohara:2021smr,Ashikawa:2024njc}, the Binder-cumulant analysis of the heavy-quark QCD CP has been performed on four-dimensional Euclidian lattices based on the HPE. In these studies, the gauge configurations are generated for the action $S_{\rm g+LO}=S_{\rm g}+S_{\rm LO}$ and the effects $S_{\rm NLO}$ are exactly taken into account by the reweighting. In Ref.~\cite{Ashikawa:2024njc}, effects of yet higher-order terms in the HPE are incorporated effectively via the method proposed in Ref.~\cite{Wakabayashi:2021eye}. We remark that this prescription enables us to perform Monte Carlo simulations quite effectively. In fact, about $10^6$ measurements have been performed in Refs.~\cite{Kiyohara:2021smr,Ashikawa:2024njc} up to the aspect ratio $LT=N_{\rm s}/N_{\rm t}=18$.
In Fig.~\ref{fig:B4}, we show the result of the Binder cumulant of the Polyakov loop, $B_4^\Omega$, as a function of $\lambda= 2^{N_t+2} N_{\rm f} N_{\rm c} \kappa^{N_t}$ around the CP at $N_t=4$~\cite{Kiyohara:2021smr}, where the statistical errors estimated by the jackknife method are shown by the shaded bands. The figure shows $B_4^\Omega$ in Eq.~\eqref{eq:B4} for aspect ratios $LT=N_\textrm{s}/N_\textrm{t}=8$, 9, 10, and 12.
The CP is given by the intersection point of these results. From the fit analysis, we obtain its location as $(\beta_{\rm c},\kappa_{\rm c})=(5.68453(22),0.0602(4))$, where $\beta$ is the inverse coupling.

\begin{figure}
    \centering
\includegraphics[width=0.8\linewidth]{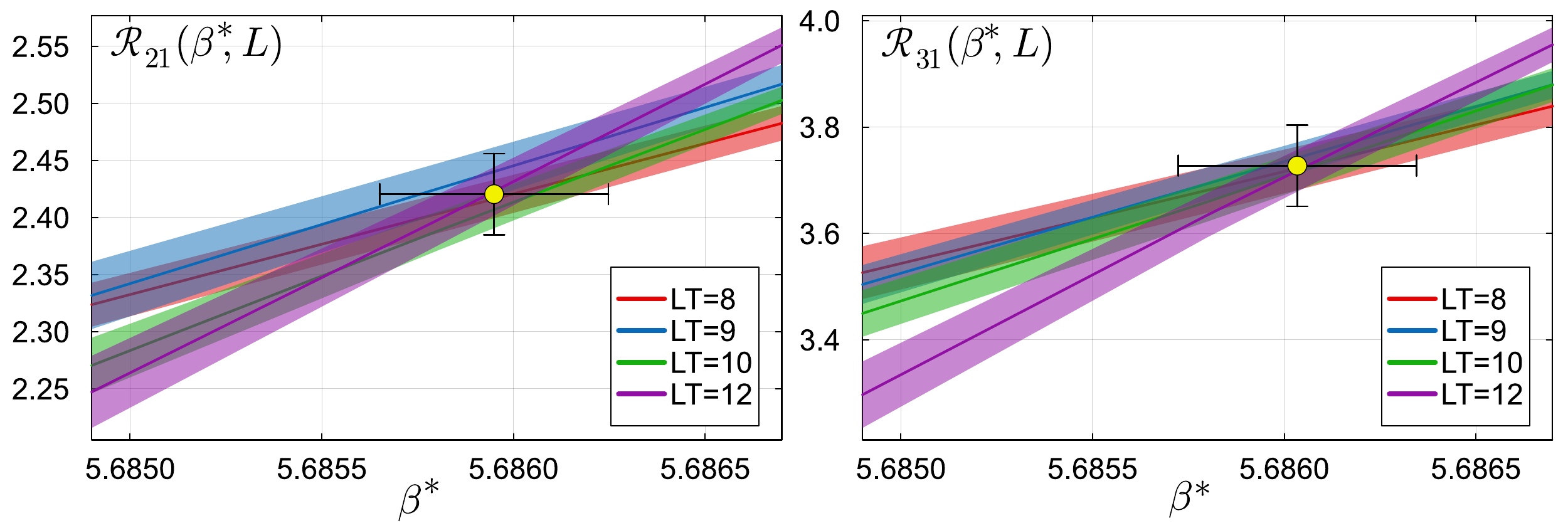}
    \caption{LYZR ${\cal R}_{21}(\tau,L)$ (left) and ${\cal R}_{31}(\tau,L)$ (right) in heavy-quark QCD at $N_t=4$ for various aspects ratios $LT=N_s/N_t$. The circle markers are the results of the FSS fits.}
    \label{fig:HQ_R}
\end{figure}

To study the behavior of LYZR around the CP in heavy-quark QCD, we performed the analysis of the LYZ on the same gauge configurations as Ref.~\cite{Kiyohara:2021smr}.
We show our results of the LYZR ${\cal R}_{21}$ and ${\cal R}_{31}$ in
Fig.~\ref{fig:HQ_R} as functions of $\beta^*=\beta+16N_cN_f\kappa^4$. 
The figure shows that ${\cal R}_{n1}$ at various $LT$ intersect with one another approximately at a common intersection point.
To determine the location of the CP, we ignore the second bracket of Eq.~(\ref{eq:Rlinear2}) and perform the FSS fits, ${\cal R}_{n1}(\beta^*,LT)=r+c(\beta^*-\beta^*_\textrm{c})(LT)^{y_t}$, for $LT\ge9$, where $r$, $c$, $\beta^*_\textrm{c}$, and $y_t$ are the fitting parameters.
The fit results are shown by the open circles in Fig.~\ref{fig:HQ_R}. These results give $\beta^*_\textrm{c} = 5.68595(30)$ and $5.68603(31)$ for $\mathcal{R}_{21}$ and $\mathcal{R}_{31}$, respectively, which are are consistent with the value obtained by the Binder-cumulant analysis, $\beta^*_{\rm c}= 5.68578(22)$~\cite{Kiyohara:2021smr}. 

\section{Summary}

In this proceeding, we proposed a novel method to investigate a CP in general systems using the ratios of LYZ in numerical simulations. 
This method, the LYZR method, is superior to the Binder-cumulant analysis in faster suppression of the finite-size effects in general systems. 
We then applied the method to an investigation of the CP in heavy-quark QCD. Our results of the Monte Carlo simulations show that the location of the CP is successfully determined by the LYZR method. A similar result is obtained on the 3-state Potts model~\cite{Wada:2024qsk}. Since these results confirm the validity of the LYZR method in practical numerical analyses, it is an interesting future study to apply the method to the analyses of CPs in other models, especially $2+1$-flavors QCD at physical-quark mass~\cite{Schmidt:2025ftp,Adam:2025pii}.

\subsection*{Acknowledgements}
We thank Shinji Ejiri for discussions in the early stage of this study. 
This work was supported in part by JSPS KAKENHI (Nos.~JP22K03593, JP22K03619, JP23H04507, JP24K07049), the Center for Gravitational Physics and Quantum Information (CGPQI) at Yukawa Institute for Theoretical Physics, the Research proposal-based use at the Cybermedia Center, Osaka University, and the Multidisciplinary Cooperative Research Program of the Center for Computational Sciences, University of Tsukuba.

\bibliographystyle{JHEP}
\bibliography{ref}

\end{document}